\documentclass[12pt]{article}
\newcommand{\be}{\begin{equation}}
\newcommand{\ee}{\end{equation}}
\newcommand{\beas}{\begin{eqnarray*}}
\newcommand{\eeas}{\end{eqnarray*}}
\newcommand{\bea}{\begin{eqnarray}}
\newcommand{\eea}{\end{eqnarray}}
\newcommand{\ba}{\begin{array}}
\newcommand{\ea}{\end{array}}
\newcommand{\nn}{\nonumber}

\newcommand{\al}{\alpha}
\newcommand{\g}{\gamma}
\newcommand{\de}{\delta}

\begin{document}
\title{
{\bf Space-time revisited}
}
\author{
Yu.\ F.\ Pirogov\\
{\it Theory Division, Institute for High Energy Physics,  Protvino,}\\
{\it  RU-142281 Moscow Region, Russia}
}
\date{}
\maketitle
\abstract{\noindent 
The metric space-time is revised as a priori existing. It is
substituted by the world continuum
endowed  only with the affine connection. The metric, accompanied by
the tensor Goldstone boson, is to emerge during  the spontaneous
breaking of the global affine symmetry. Implications for 
gravity and the Universe are indicated.}

\section{Introduction}

According to the present-day paradigm, the physical sciences start
with the  space-time possessing metric as the
primordial structure. I~propose to go beyond this paradigm and
substitute the metric space-time, at the underlying level, by  the
world continuum which possesses only the affine connection.
The metric is to appear at the effective level during
the world structure formation. In other words, the space time
is to change its status from  a~priori existing to emerging. 
Ultimately, this approach  results in the nonlinear model
$GL(4,R)/SO(1,3)$ for
the gravity, with the graviton as the tensor Goldstone boson
corresponding to the spontaneously broken global affine symmetry
(in detail, see ref.~\cite{Pirogov}). This is the
further development of the Goldstone approach to
gravity~\cite{Salam,Ogievetsky1}.

\section{Affine symmetry}\label{aff}

\paragraph{Affine connection}

Assume that the forebear of the space-time
is the world continuum equipped only with the affine
connection. Let~$x^\mu$, $\mu=0,\dots,3$  be the world coordinates.
There being, prior to  metric,
no  partition  of the continuum onto the space and time, the
index~$0$ has yet no particular meaning.
In ignorance of the underlying ``dynamics'',
consider all the structures related to the
continuum as the background ones. Let
$\bar\psi^\lambda{}_{\mu\nu}(x)$ be the  background
affine connection and  let $\bar\xi^{\al}$ be the coordinates where
the connection have a particular, to be defined,
form ${\bar \psi}{}^{\g}{}_{\al\beta}(\bar\xi)$.\footnote{The bar sign
refers to the background. The indices
$\al$, $\beta$, etc,  refer to  the special coordinates, while
$\lambda$, $\mu$, etc, to the world ones.}  For
reason not to be discussed, let the antisymmetric part of the
background connection (the torsion)  be absent identically.
As for the symmetric part,  one is free to choose the  special
coordinates to make the physics description as clear as possible.
 
So, let $P$ be a fixed but otherwise arbitrary point  with the world
coordinates~$X^\mu$. One can nullify the symmetric part of the
connection in this point by adjusting the proper coordinates
$\bar\xi^{\al}(x,X)$. In the vicinity of~$P$, the connection becomes:
\be\label{eq:connection}
{\bar \psi}^{\g}{}_{\al\beta}(\bar\xi)= \frac{1}{2}\,
{\bar \rho}^{\g}{}_{\al\delta\beta}(\bar\Xi)\,
(\bar\xi-\bar\Xi)^{\de} + {\cal O}((\bar\xi-\bar\Xi)^2),
\ee
with ${\bar \rho}^{\g}{}_{\al\delta\beta}(\bar\Xi)$
being the background  curvature tensor in the 
point~$P$ and $\bar\Xi\equiv \bar\xi^{\al}(X,X)$.
Let us consider the whole set of the  coordinates
with the proper\-ty $\bar \psi^{\al}{}_{\beta\g}\vert_P=0$.
The allowed group of transformations of such 
coordinates is the inhomogeneous general linear group
$IGL(4,R)=T_4 \times GL(4,R)$ (the affine one):
\be\label{eq:affine}
(A,a)\ : \ \bar\xi^{\al} \to
\bar\xi'^{\al}=A^{\al}{}_{\beta}
\bar\xi^{\beta}+a^{\al},
\ee
with $A$ being an arbitrary nondegenerate matrix.
Under these, and only under these transformations, the affine
connection in the point~$P$ remains to be zero. The group is
the global one in the sense that it transforms the local, i.e., the
point~$P$ related coordinates in the global manner,  i.e., for all the
continuum at once.
The respective coordinates will be called the local affine
ones.\footnote{The term ``local'' will be omited for short.} In these
coordinates, the continu\-um in a neighbourhood of the  point
is approximated by  the affinely flat manifold. In particular,  the
underlying covariant derivative in the affine coordinates in the
point~$P$ coincides with the partial derivative.

\paragraph{Beyond the special relativity}

According to the special relativi\-ty, the present-day physical laws
are invariant relative to the choice of the inertial coordinates, with
the space-time symmetry being the Poincare one.
Postulate the principle of  {\it the extended relativi\-ty},  stating
the invariance  relative to the choice of the affine coordinates.
The physics invariance symmetry extends now to the affine group. The
latter is 20-parametric and supplement the 10-parameter Poincare group
$ISO(1,3)$ with the ten special affine transformations.
There being known  no exact
affine symmetry,  the latter should be broken to the Poincare symmetry
in transition from the underlying level to the effective one.

\paragraph{Metric and symmetry breaking}

Assume that the affine symmetry breaking is achieved due to the
spontaneous emergence  of the background
metric $\bar\varphi_{\mu\nu}(x)$ in the world
continuum. The metric is assumed to have the Minkowskian signature and
to look in the affine coordinates as:
\be\label{eq:metric}
\bar \varphi_{\al\beta}(\bar\xi)=
\bar\eta_{\al\beta}-\frac{1}{2}
\bar
\rho_{\g\al\delta\beta}(\bar\Xi)\,
(\bar\xi-\bar\Xi)^{\g}(\bar\xi-\bar\Xi)^{\de}+{\cal
O}((\bar\xi-\bar\Xi)^3).
\ee
Here one puts   $\bar\eta_{\al\beta}\equiv \bar
\varphi_{\al\beta}(\bar\Xi)$ and 
$\bar \rho_{\g\al\delta\beta}(\bar\Xi)=
\bar \eta_{\g\de}\,
\bar \rho^{\de}{}_{\al\delta\beta}(\bar\Xi)$.
The metric~(\ref{eq:metric}) is such that the Christoffel
connection $\bar\chi^\g{}_{\al\beta}(\varphi)$, determined by the
metric, matches with the  affine
connection $\bar\psi^\g{}_{\al\beta}$ in the sense that the
connections coincide locally, up to the first derivative:
$\bar\chi^\g{}_{\al\beta}=\bar\psi^\g{}_{\al\beta}+{\cal
O}((\bar\xi-\bar\Xi)^2)$. 
This is reminiscent of the well-known fact that the metric in
the Riemannian manifold may be  approximated locally, up to the
first derivative, by the Euclidean metric. In the wake of the
background  metric, there appears the (yet preliminary)
partition of the world continuum onto the space and time. 
 
Under the  affine symmetry, the  background metric
ceases to be invariant. But it still possesses an
invariance subgroup. Viz., without any loss of
generality, one can choose among the  affine
coordinates the particular ones with $\bar \eta_{\al\beta}$
being in the Minkowskian form $\eta=\mbox{diag}\,(1,-1,-1,-1)$. The
respective coordinates will be called the background inertial
ones. They are to be distinguished from the effective inertial ones
(see later on). Under the affine transformations, one has 
\be\label{eq:eta_lin}
(A,a)\ : \ \eta\to \eta'=A^{-1T}  \eta A^{-1}\ne\eta, 
\ee
whereas  the Lorentz transformations $A=\Lambda$ still leave $\eta$
invariant.
It follows that the subgroup of invariance of $\eta$ is the Poincare
group $ISO(1,3)\in IGL(4,R)$, the translation subgroup being intact. 
Thus under the appearance of the metric, the $GL(4,R)$ group is
broken spontaneously to  the residual Lorentz one
\be
GL(4,R)\stackrel{M_A\ }{\longrightarrow} SO(1,3).
\ee 
For the symmetry breaking scale $M_A$, one expects a priori
$M_A\sim M_{Pl}$,  with $M_{Pl}$ being the Planck
mass.  The relation between the scales will be discussed later on.

\section{Gravity}\label{AGB}

\paragraph{Affine Goldstone boson}

Let $\bar\xi^{\al}$ be the background inertial coordinates
adjusted to the  point~$P$. Attach to this  point 
the auxiliary linear space~$T$, the tangent space in the point.
By definition, $T$ is isomorphous to the Minkowski
space-time. The tangent space is the
structure space of the theory, whereupon the
realizations/representations of the physics space-time
symmetries, the affine and the Poincare ones, are defined. 
Introduce in  $T$ the coordinates $\xi^{\al}$,
the counterpart of the background inertial
coordinates $\bar\xi^{\al}$ in the space-time. 
By construction, the connection in the tangent space is zero
identically. For the connection in the space-time
in the the point~$P$ to be zero, too,  the coordinates are
to be related as 
$\xi^{\al}= \bar\xi^{\al} +{\cal O}((\bar\xi-\bar\Xi)^3)$.

Due to the spontaneous breaking, $GL(4,R)$ should be
realized in the nonlinear manner~\cite{NL}, with the
nonlinearity scale $M_A$, the Lorentz symmetry being still realized
linearly. The spinor representations of the latter correspond to the
matter fields, as usually. In this, the finite dimensional
spinors appear only at the level of $SO(1,3)$.
The broken part $GL(4,R)/SO(1,3)$ should be
realized in the Nambu-Goldstone mode. Accompanying  the spontaneous
emergence of the metric, there should appear the 10-component
Goldstone boson which corresponds to  the ten  generators of the
broken affine transformations. 

According to ref.~\cite{NL}, the  nonlinear 
realization  of the symmetry~$G$ spontaneously
broken to the symmetry $H\subset G$ can be built on the
quotient space $K=G/H$, the residual subgroup~$H$ serving as the
classification group. One is interested in the  
pattern $GL(4,R)/SO(1,3)$, with the quotient space consisting of all
the broken affine transformations.
Let $\kappa(\xi)\in K$ be the coset-function on the tangent space. To
restrict $\kappa$ by the quotient space, one should impose on the
representative group element some auxiliary condition, eliminating
explicitly the extra degrees of freedom.
Under the arbitrary affine transformation
$\xi \to  \xi'=A\xi+ a$, the coset is to transform~as
\be\label{eq: k}
(A,a)\ :\  \kappa(\xi) \to \ \kappa'(\xi')=
A \kappa(\xi)\Lambda^{-1},
\ee
where $\Lambda(\kappa,A)$ is the appropriate element of the residual
group, here the Lorentz one. This makes the transformed element
compatible with the auxiliary condition.
In the same time, by the very  construction, the Minkowskian~$\eta$
stays invariant under the nonlinear realization: 
\be\label{eq:eta}
(A,a)\ : \ 
\eta\to\eta'=\Lambda^{-1T}\eta\Lambda^{-1}=\eta
\ee
(in distinction with the linear representation
eq.~(\ref{eq:eta_lin})).

Otherwise, one can abandon any auxiliary condition
extending the affine symmetry by the hidden local symmetry 
$\hat H\simeq H$. The extra 
Goldstone degrees of freedom are  now unphysical due to the gauge 
transformations~$\Lambda(\xi)$. This is the linearization of
the nonlinear model, with the proper gauge boson $v_{ab\g}$ 
being expressed, due to the equation on motion, through 
$\hat\kappa^a_\al$ and its derivatives. With this in mind, the abrupt
expressions entirely in terms of $\hat\kappa^a_\al$ and its
derivatives are used in what follows. The versions
differ by the higher order corrections.

In the tangent space, one should distinguish now 
two types  of indices: the Lorentz ones, acted on by the
local Lorentz transformations~$\Lambda(\xi)$, and the affine ones,
acted on by the global affine transformations~$A$. Designate the
Lorentz indices as $a, b$, etc,
while the affine ones as before $\al, \beta$, etc. 
The Lorentz indices are manipulated by
means of the Minkowskian $\eta_{ab}$ (respectively, $\eta^{ab}$).
The Goldstone field is represented by the arbitrary $4\times 4$
matrix ${\hat \kappa}^\al_a$ (respectively,  
${\hat \kappa}^{-1}{}^a_\al$) which  transforms
similar to eq.~(\ref{eq: k}) but with arbitrary
$\Lambda(\xi)$.\footnote{The
hat sign refers to the hidden local Lorentz symmetry.}

\paragraph{Matter and radiation}

For the matter fields $\phi$ one puts
\be\label{eq:phi}
\phi(\xi)\to\phi'(\xi')=\hat\rho_\phi(\Lambda)\phi(\xi),
\ee
with $\hat\rho_\phi$ taken in the proper Lorentz representations. 
As for the gauge bosons, they
constitutes one more separate kind of fields, the radiation. 
By definition, the gauge fields  $V_{\al}$ transform under
$A$ linearly as the derivative $\partial_{\al}\equiv\partial/\partial
\xi^{\al}$. The modified fields
$\hat V_a \equiv {\hat\kappa}^\al_a V_{\al}$
transform as the Lorentz vectors 
\be\label{eq:barV}
\hat V(\xi)\to  \hat V'(\xi')=\Lambda^{-1T}  \hat V(\xi)
\ee
and are to be used in the model building.

\paragraph{Nonlinear model}\label{MN}

To explicitly account for the residual symmetry it is convenient to
start with the objects transforming only under the latter symmetry.
Clearly, any nontrivial combinations of ${\hat\kappa}$
and ${\hat\kappa}^{-1}$ alone transform explicitly under $A$. Thus the
derivative terms are inevitable. To describe the latter ones, 
introduce the Cartan one-form chosen  as follows: 
\be\label{eq:Cartan}
{\hat\omega}=\eta {\hat\kappa}^{-1}d {\hat\kappa}.
\ee
The one-form transforms as the Lorentz quantity:
\be\label{eq:Delta_trans}
{\hat\omega}(\xi) \to
{\hat\omega}'(\xi')=\Lambda^{-1T}{\hat\omega}(\xi)\Lambda^{-1}
+ \Lambda^{-1T}\eta d\Lambda^{-1}.
\ee

In the component notation, the one-form looks like
${\hat\omega}_{ab}$. Decompose it into  the  symmetric and
antisymmetric
parts ${\hat\omega}^\pm_{ab}$, respectively:
\be
{\hat\omega}_{ab}\equiv \sum_\pm {\hat\omega}^\pm_{ab}=\sum_\pm 
[\eta {\hat\kappa}^{-1}d{\hat\kappa}]^\pm_{ab}.
\ee
One can see that ${\hat\omega}^\pm_{ab}$ transform independently as
\be\label{eq:MC}
{\hat\omega}^\pm(\xi)\to 
{\hat\omega}'^{\pm}(\xi')=\Lambda^{-1T}\hat\omega^{\pm}(\xi) 
\Lambda^{-1} +\delta^\pm,
\ee
where
\bea
\delta^-&=&\Lambda^{-1T}\eta d\Lambda^{-1},\nn\\
\delta^+&=&0.
\eea
For the  derivative of the one-form one gets:
\be
\label{eq:B}
\hat\omega^{\pm}_{abc}=
{\hat\kappa}^{\g}_c \,{\hat\omega}^\pm_{ab}/ \partial\xi^{\g}= [\eta
{\hat\kappa}^{-1}
\hat\partial_c^{\phantom \pm}\! {\hat\kappa}]^{\pm}_{ab},
\ee
where  
$\hat\partial_c \equiv
{\hat\kappa}^\g_c \partial_{\g}= {\hat\kappa}^\g_c \partial/\partial
\xi^{\g}$
is the effective partial derivative. 
Transforming inhomogeneously,  $\hat\omega^{-}_{abc}$ could be used as
the minimal connection for the nonlinear realization.

The transformation properties of the nonlinear 
covariant derivative are not changed if one adds to the
above minimal connection the properly modified
terms $\hat\omega^+_{abc}$, the latter ones transforming
homogeneously. For consistency reason (see later on),
choose for the nonminimal connection the following special
combination: 
\be\label{eq:c}
\hat\omega_{abc}=\hat\omega^-_{abc} +
\hat\omega^{+}_{cab} -\hat\omega^{+}_{cba}.
\ee
By means of this connection, one can define the nonlinear
derivatives of the matter fields $\hat D_a\phi$, the gauge strength
$\hat F_{ab}$, as well as the field strength for
affine Goldstone boson $\hat R_{abcd}$ and its contraction $\hat
R\equiv\eta^{ab} \eta^{cd}\hat R_{abcd}$.

The above objects can serve
as the building blocks for the  nonlinear model $GL(4,R)/SO(1,3)$ 
in the tangent space. Postulate {\it the
equivalence principle} in the sense that the  tangent space Lagrangian
should not depend explicitly on the background parameter-functions
$\bar\rho^d{}_{abc}$ (cf.\ eq.~(\ref{eq:connection})).
Thus, the Lagrangian  may  be written as the general  Lorentz  (and,
thus, affine) invariant built of $\hat R$, 
$\hat F_{ab}$, $\hat D_a \phi$ and~$\phi$. As usually, one
restricts himself by the terms containing two derivatives at the most.

Once such a Lagrangian is built, one can rewrite it by means of
${\hat\kappa}^\al_a$ and
${\hat\kappa}^{-1}{}^a_\al$ in terms of the
affine quantities. This makes explicit the  geometrical
structure of the theory and  relates the latter with the gravity. 
Under the above choice for the
nonlinear connection, the Lagrangian for the affine Goldstone boson,
radiation and matter becomes
\be\label{eq:L}
L= c_g M_A^2  R (\g_{\al\beta}) +
{L}_{r}(F_{\al\beta})
+  {L}_m (D_{\al}\phi,\phi ).
\ee 
Here 
\be\label{eq:gamma'}
\g_{\al\beta}={\hat\kappa}^{-1}{}^a_{\al}\eta_{ab}
{\hat\kappa}^{-1}{}^b_{\beta}
\ee
transforms as the affine tensor 
\be\label{eq:g_aff}
(A,a)\ : \  \g_{\al\beta}\to \g'_{\al\beta}=A^{-1}{}^\g{}_\al
\g_{\g\delta}A^{-1}{}^\delta{}_{\beta}.
\ee
It proves that $ R(\g_{\al\beta})=\hat R(\hat\omega_{abc})$ can
be expressed as the contraction $R=R^{\al\beta}{}_{\al\beta}$ of the
tensor $R^\g{}_{\al\delta\beta}\equiv \eta^{\g\g'}
{\hat\kappa}^{-1}{}^a_\al
{\hat\kappa}^{-1}{}^b_\beta   {\hat\kappa}^{-1}{}^d_\de
{\hat\kappa}^{-1}{}_{\g'}^c  \hat R_{cadb}$,
the latter in
turn being related with $\g_{\al\beta}$ as the Riemann-Christoffel
curvature tensor with the metric. In this, all the
contractions of the affine indices are understood with 
$\g_{\al\beta}$ (respectively, $\g^{\al\beta}$). 
Similarly,  $ D_{\al}\phi \equiv
{\hat\kappa}^{-1}{}_\al^a \hat D_a\phi$ look like the  covariant
derivatives of the  matter fields with the spin-connection
$\omega_{ab\g}\equiv\hat\omega_{abc}\, {\hat\kappa}^{-1}{}^{c}_{\g}$.
The gauge strength $F_{\al\beta}$ has the usual form containing the
partial derivative $\partial_\al$.

Clearly, the Lagrangian $L$ looks like the generally covariant one in
the tangent space considered as the  Riemannian manifold with 
the effective\footnote{The term ``effective'' will be omitted for
short, while that ``background'' will,
in contrast, be retained.}  metric~$\g_{\al\beta}$,  the
Riemann-Christoffel curvature
$R^\g{}_{\al\delta\beta}$, the Ricci curvature~$R_{\al\beta}$, the
Ricci scalar~$R$, the spin-connection $\omega_{ab\g}$  and the tetrad
${\hat\kappa}^{-1}{}^a_\al$  (the inverse one ${\hat\kappa}^\al_a$). 
This is in no way accidental. Namely, as it is  shown in
ref.~\cite{Ogievetsky1}, under the special choice  of the
nonlinear connection eq.~(\ref{eq:c}),
the Lagrangian becomes conformally invariant, too. 
Further, according to ref.~\cite{Ogievetsky2}, the theory which is
invariant both under the conformal
symmetry and  the global affine one is generally covariant.
After the choice of the metric, this imposes the Riemannian structure
onto the tangent space. 
Precisely the last property justifies the above special choice for the
nonlinear connection. The affine
Goldstone boson proves to be the graviton in disguise.

\paragraph{General Relativity and beyond}

The preceding construction  referred to 
the tangent space~$T$ in the given point~$P$.  
Accept the so defined  Lagrangian  as that for the  space-time,
being valid in the background  inertial coordinates in the
infinitesimal neighbourhood of the point.
After  multiplying the Lagrangian by the
generally covariant volume element $(-\g)^{1/2}\,d^4
\bar\Xi$\,, with  $\g\equiv \mbox{det}\g_{\al\beta}$, one gets
the infinitesimal contribution into the action in the given
coordinates.

The relation between the  background inertial  and  world coordinates
is achieved by means of the background  tetrad  
$\bar e^{\al}_\mu(X)$. Now, introduce the effective tetrad related
with the background one as 
\be\label{eq:bare}
e_\mu^a(X)={\hat\kappa}^{-1}{}^a_{\al}(X)\,\bar e_\mu^{\al}(X).
\ee
The effective tetrad transforms  as  the Lorentz vector:
\be\label{eq:Le}
e_\mu(X)\to e'_\mu(X)=\Lambda(X)\, e_\mu(X).
\ee
Due to the local Lorentz transformations $\Lambda(X)$, one can 
eliminate six components out of~$e_\mu^a$, the latter having 
thus ten physical components. In this terms, the effective metric in
the world coordinates is
\be
g_{\mu\nu}\equiv  \bar e^\al_\mu \g_{\al\beta}\bar
e^\beta_\nu= e^a_\mu \eta_{ab}e^b_\nu.
\ee
In other words, the tetrad  $e^a_\mu$ defines the effective inertial
coordinates. Physically, eq.~(\ref{eq:bare}) describes the
disorientation of the effective inertial and background inertial
frames depending on the distribution of the affine Goldstone boson. 

By means  of $e^a_\mu$, the tangent space quantities result in
the world coordinates in the usual
expressions of the Riemannian geometry containing  metric
$g_{\mu\nu}$ and  spin-connection $\omega_{ab\mu}$. One
gets for the total action:
\be\label{eq:L_AGB'}
I=\int \left(
\frac{1}{2}M_{Pl}^2 R(g_{\mu\nu})
+\,L_r( F_{\mu\nu})
+L_m({D}_\mu \phi,\phi)\right)
(- g)^{1/2}\,d^4 X,
\ee
with $ g\equiv \mbox{det}\, g_{\mu\nu}$. 
Note that due to the weight factor $\sqrt{-g}$, the
affine Goldstone boson  enters the action also with the
derivativeless couplings. 
Finally, one arrives at the General Relativity (GR) equation of motion
for gravity:
\be\label{eq:Einstein'}
{R}_{\mu\nu} -\frac{1}{2}R\, g_{\mu\nu}
=M_{Pl}^{-2}T_{\mu\nu}.
\ee
Here $T_{\mu\nu}$ is the 
energy-momentum tensor of the radiation and  matter.

In the above, the constants
are such that $c_g M_A^2=1/2\,M^2_{Pl}\equiv 1/(16\pi
G_N)$, with $G_N$ being the Newton's constant. 
Superficially, the (effective) Riemannian geometry  is valid
at all the space-time intervals. Nevertheless, its accuracy  worsen
at the smaller and smaller intervals, requiring more and more terms in
the decomposition over the ratio of the energy to the symmetry
breaking scale~$M_A$, as it should be for the effective theory.
Thus, the scale~$M_A$ (or rather, the Planck mass $M_{Pl}$) is  the
inverse minimal length in the nature. 
 
In the GR, after fixing the Lagrangian  the theory
becomes unique, independent of  the choice of the coordinates.
Under extension of the tangent space Lagrangian beyond the general
covariance,  the theory in the space-time ceases to be generally
covariant and thus  unique. It
depends not only on the Lagrangian but
on the choice of the coordinates. Relative to  the general
coordinate transformations, the obtained GR extensions divide into
the inequivalent classes, each of which  is  characterised by the
particular set of the background parameter-functions.  A~priori,
no one of the sets is preferable. Which one is suitable (if any),
should be determined by observations. 
Each class consists of the equivalent extensions related by the
residual covariance group. 
Among the inequivalent extensions, there appears the natural hierarchy
according to whether the affine symmetry  is violated or not. For
details, see ref.~\cite{Pirogov}.

\paragraph{The Universe and beyond}\label{ESV}

Let the formation of the Universe be the result of the
actual transition between the two phases of the continuum, the
affinely connected and metric ones. 
This transition is thus  {\it the ``Grand Bang''}, the origin of the
Universe and the very space-time. 
There is conceivable the appearance (as well as 
disappearance and coalescence) of the various bubbles of the metric
phase inside the affinely connected  one (and v.v.). 
These  bubbles are to be associated with the multiple universes.
One of the latter ones happens to be ours.
Hopefully, this may shed light on  the long-standing problem of the
fine tuning of our Universe. 

\section{Conclusion}

In conclusion, the new physics paradigm  realizes
consistently the approach to gravity as the Goldstone phenomenon.
The theory constructed proceeds, in essence, from two basic
symmetries: the spontaneously broken global affine symmetry and the
general covariance. 
The theory embodies the GR as the lowest approximation. 
Its distinction with the GR are twofold. At the effective level, the
theory predicts the natural hierarchy of the  GR extensions depending
on the mode of the affine symmetry. At the underlying level, it
presents the new physics comprehension of the gravity and  the
Universe.

\paragraph{Acknowledgement} 
I am grateful to the  Organizing Committee of the 13th International
Seminar {\it ``Quarks--2004''} for support.

\end{document}